\documentclass[pre,floatfix,longbibliography,superscriptaddress,twocolumn,final,notitlepage,nofootinbib]{revtex4-1}
\usepackage{tabularx}
\usepackage[caption=false]{subfig}
\usepackage{graphicx}
\graphicspath{{./figures/}} % Directory in which figures are stored
\usepackage{amsfonts,amsmath,amssymb}
\usepackage{comment}
\usepackage{lineno}
\usepackage{hyperref}
\usepackage[explicit]{titlesec}
\usepackage{footmisc}
\newcommand{\be}{\begin{equation}}
\newcommand{\ee}{\end{equation}} 
\newcommand{\bea}{\begin{eqnarray}}
\newcommand{\eea}{\end{eqnarray}}

\usepackage[T1]{fontenc}
\usepackage[utf8]{inputenc}

\usepackage{tgtermes}
\usepackage{amsmath}
\usepackage{scalefnt,letltxmacro}
\LetLtxMacro{\oldtextsc}{\textsc}
\renewcommand{\textsc}[1]{\oldtextsc{\scalefont{1.10}#1}}
\usepackage[scaled=0.92]{PTSans}
\usepackage{inconsolata}
\usepackage{mathbbol}

\usepackage[usenames,dvipsnames]{xcolor}
\definecolor{shadecolor}{gray}{0.9}

\usepackage[english]{babel}
\usepackage{afterpage}
\usepackage{framed}
\usepackage{nicefrac}

{\endMakeFramed}

\DeclareRobustCommand{\parhead}[1]{\textit{#1}~}

\usepackage{booktabs}
\usepackage{multirow}
\usepackage{longtable}
\usepackage{etoolbox,siunitx}
\robustify\bfseries
\usepackage{natbib}
\usepackage{bibunits}

\usepackage[algoruled,algo2e,linesnumbered]{algorithm2e}
\usepackage{listings}
\usepackage{fancyvrb}
\fvset{fontsize=\normalsize}
\usepackage{algorithm}
\usepackage{algorithmic}

\usepackage[nameinlink]{cleveref}

\usepackage[acronym,smallcaps,nowarn]{glossaries}
\glsdisablehyper{}
\usepackage{listings}
\lstdefinestyle{mystyle}{
    commentstyle=\color{OliveGreen},
    numberstyle=\tiny\color{black!60},
    stringstyle=\color{BrickRed},
    basicstyle=\ttfamily\scriptsize,
    breakatwhitespace=false,
    breaklines=true,
    captionpos=b,
    keepspaces=true,
    numbers=none,
    numbersep=5pt,
    showspaces=false,
    showstringspaces=false,
    showtabs=false,
    tabsize=2
}
\usepackage[charter]{mathdesign}  % for the Expected value symbol of capital E

\usepackage{microtype}

\usepackage{amsthm}
\theoremstyle{definition}
 
\theoremstyle{notation}
 
\theoremstyle{theorem}
\newtheorem{theorem}{Theorem}
\usepackage{amsthm}
\theoremstyle{definition}
\newtheorem{lemma}{Lemma}

\hypersetup{
 bookmarks=true, % show bookmarks bar?
 unicode=false, % non-Latin characters in Acrobat�s bookmarks
 pdftoolbar=true, % show Acrobat�s toolbar?
 pdfmenubar=true, % show Acrobat�s menu?
 pdffitwindow=false, % window fit to page when opened
 pdfstartview={FitH}, % fits the width of the page to the window
 pdftitle={}, % title
 pdfauthor={}, % author
 pdfcreator={}, % creator of the document
 pdfproducer={}, % producer of the document
 pdfkeywords={} {} {}, % list of keywords
 pdfnewwindow=true, % links in new window
 colorlinks=true, % false: boxed links; true: colored links
 linkcolor=red, % color of internal links
 citecolor=ForestGreen, % color of links to bibliography
 filecolor=magenta, % color of file links
 urlcolor=cyan % color of external links
}
\usepackage[textsize=tiny,color=SkyBlue]{todonotes}

\usepackage{tabularx}
\usepackage{adjustbox}
\usepackage{url} 

\graphicspath{{./figures/}} % Directory in which figures are stored
\usepackage{booktabs} % for professional tables

\usepackage{xcolor,colortbl}
\usepackage{hyperref}

\usepackage{times}
\usepackage{tcolorbox}
\usepackage{enumitem}

\begin{document}

\title{Sampling on networks: estimating eigenvector centrality on incomplete graphs}

\author{Nicol\`{o} Ruggeri}
\email{nicolo.ruggeri@tuebingen.mpg.de}
\address{ Max Planck Institute for Intelligent Systems, Max-Planck ring 4, Tübingen, 72076, Germany}
\address{ Universit\`{a}  di Padova, Department of Mathematics ``Tullio Levi-Civita'', via Trieste 63, Padova, 35100,  Italy}

\author{Caterina De Bacco}
\email{caterina.debacco@tuebingen.mpg.de}
\address{ Max Planck Institute for Intelligent Systems, Max-Planck ring 4, Tübingen, 72076, Germany}

% Optional adjustment to line up main text (after abstract) of first page with line numbers, when using both lineno and twocolumn options.
% You should only change this length when you've finalised the article contents.
%\verticaladjustment{-2pt}

%%%%%%%%%%%%%%%%%%%%%%%%%%%%%%%%%%%%%%%%%%%%%%%%%%%%%%%%%%
\begin{abstract}
  We develop a new sampling method to estimate eigenvector centrality on incomplete networks.
    Our goal is to estimate this global centrality measure having at disposal a limited amount of data. This is the case in many real-world scenarios where data collection is expensive, the network is too big for data storage capacity or only partial information is available.
 The sampling algorithm is theoretically grounded by results derived from spectral approximation theory.
 We studied the problem on both synthetic and real data and tested the performance comparing with traditional methods, such as random walk and uniform sampling.
We show that approximations obtained from such methods are not always reliable and that our algorithm, while preserving computational scalability, improves performance under different error measures.
\end{abstract}

\maketitle

%%%%%%%%%% Introduction

\section{Introduction}
\label{introduction}

One key task in inference in networks is to evaluate quantities when network information is limited as when the available network is only a subsample of the full one.
Such evaluations impacts problems like modeling dynamical processes \cite{de2010does,sadikov2011correcting}, network statistics estimation \cite{leskovec2006sampling}, data compression \cite{adler2001towards} and survey design \cite{frank2005network}. 
The sampling procedure used to collect network information usually influences these estimates \cite{han2005effect,lee2006statistical,kossinets2006effects}, it is therefore important to choose a sampling algorithm properly based on what is the quantity one seeks to estimate.
The challenge is to provide a reliable estimate while keeping the sampling algorithm scalable. 
While several sampling strategies can be found in the literature \cite{leskovec2006sampling}, these are usually empirically-driven and lack of analytical groundings . 

Here we take a theoretically motivated approach to approximate in-sample eigenvector centrality \cite{bonacich1972factoring}, a network centrality measure that is widely used across domains and that is highly sensitive to missing information compared to other standard centrality measures \cite{costenbader2003stability}. 
To set up the model, we minimize a proper distance between the centrality vector estimated from the subsample and the ground truth one. This allows to preserve the relative ranking of nodes inside the sample, ordered by their estimated centralities.

To tackle this problem, we develop a novel online sampling algorithm. The idea is that a proper sampling procedure should be capable of selecting nodes in an online fashion. We argue that, in order to achieve scalability, only local information should be used. This is challenging, as eigenvector centrality is considered a \textit{global} centrality measure, i.e. its calculation requires global network information. 

Using results derived from numerical methods for eigenvalues approximation \cite{saad}, our sampling algorithm provides valid centrality estimates. We derive an effective criterion based on theoretical bounds that we develop and adapt to our case. \\
Evaluation is performed on both synthetic and real data and performance is measured in terms of recovering the relative ranks of nodes inside the sample.
We found that our algorithm leads to sensible improvements for several network structures, in particular for undirected ones. 

\parhead{Related work.} The problem of reliable network sampling has been treated in the literature mainly using empirical approaches \cite{blagus2017empirical, costenbader2003stability} and focusing on recovering standard topological properties like degree distribution, diameter or clustering coefficient \cite{leskovec2006sampling,morstatter2013sample, stutzbach2009unbiased, hubler2008metropolis, stumpf2005sampling, ganguly2018estimation,antunes2018sampling}. To the best of our knowledge, sampling has not been previously applied to estimating eigenvalue centrality, despite the
vast amount of literature in this area and the number of applications that use this centrality measure.
In general, few theoretical results have been developed in sampling networks, making ours one of the few method that goes beyond heuristics or empirical reasoning. In the case of PageRank score \cite{brin1998anatomy} (which can be thought of as the eigenvector centrality for a specific class of adjacency matrices) work has been done towards estimating single nodes' scores or approximating the external information missing for reliable estimation of the scores in the sample \cite{sakakura2014improved, chen2004local}.

\section{The problem}

Consider the adjacency matrix $A$ of a network $G = (\mathcal{V}, \mathcal{E})$ of size $|\mathcal{V}|=V$, where $\mathcal{V}=\{1,\dots,V\}$ is the set of nodes and $\mathcal{E}$ the set of edges, $A_{ij} \in \mathbb{R}_{\geq 0}$ is the weight of the edge from $j$ to $i$. Notice that, while in practice many real network applications consider only discrete, and possibly binary, entries, our results are valid in general for real-valued non-negative edge weights. We define $d_{in}^{G}(i):=\sum_{j \in \mathcal{V}} A_{ij}$ the weighted in-degree of node $i$. The problem of sampling the network under a particular sampling technique can be viewed as selecting a \textit{principal submatrix} $A_m'$ of size $m\leq V$, induced by selecting a subset of nodes $\mathcal{I} \subseteq \mathcal{V}$, where $|\mathcal{I}|=m$. The subsample network is denoted as $G_{m}=(\mathcal{I},\mathcal{E}_{m})$, and $\mathcal{E}_{m}\subseteq \mathcal{E}$ is the set of edges in the subsample. 
Ideally, the subnetwork $G_{m}$ should be selected as to give a good representation of the entire network $G$. This means that network-related quantities estimated on $G_{m}$ should be a good approximation for the same quantities estimated on $G$. 
In our case we are interesting in estimating the eigenvector centrality, a popular centrality measure that is been used in several contexts \cite{bonacich1972factoring}. The centrality $u_{i}$ of node $i$ is formally obtained as:
\be\label{eqn:ec}
 \lambda \, \mu_i = \sum_{j \rightarrow i} A_{ij} \mu_j \quad,
\ee
where $\lambda$ is the maximum eigenvalue of $A$.
Formally, if we denote with $\mu \in \mathbb{R}^{V}$ and $\tilde{\mu} \in \mathbb{R}^{m}$ the principal eigenvectors for $A$ and $A_m'$ respectively and consider $\mu_{m} \in \mathbb{R}^{m}$ the restriction to the sample of $\mu$, we want to find a subsample network $G_{m}$ that minimizes the total distance $d(\mu_{m},\tilde{\mu})$, under a vectorial distance metric $d(\cdot, \cdot)$. Notice that this means that our goal is to well approximate the centrality of \textit{sampled} nodes (with $\tilde{\mu}$), rather than reconstructing the entire centrality $\mu$. Given that the eigenvector centrality is often used as a measure for ranking nodes, we will consider two error measures relevant in this application, the Spearman and Kendall correlations, see section \ref{sec:empirical}.

\subsection{A theoretically motivated sampling method}
Since we are interested in estimating eigenvectors, we use projection methods for spectral approximation. The idea is finding eigenvalues and eigenvectors by computing them on smaller subspaces and then projecting them backwards into the original (bigger) space. 
As anticipated above, in our case we are interested on better approximating only the eigenvector values relative to nodes in $G_m$, i.e. the values computed on the projected space.

In the context of projection methods, theoretical bounds for the goodness of approximation of the spectrum of $A$ have been developed \cite{saad}. 

We derive a theorem that specializes these findings to our case as follows. Denote with $P_m \in \mathbb{R}^{V,V} $ the projector onto the subspace $K_m := <e_i>_{i \in \mathcal{I}}$, where $e_i \in \mathbb{R}^V$ are the vectors of the naive basis of $\mathbb{R}^V$. In our case we are interested in approximating $P_{m}\mu$, the eigenvector centrality of the the nodes in the subsample. The zero-padding submatrix of $A$ where the rows and columns belonging to $\mathcal{I}$ are selected is defined as $A_m := P_m \,A\, P_m= \begin{pmatrix}
A_m' & 0 \\
0 & 0
\end{pmatrix}$.  Notice that $A_m \in \mathbb{R}^{V, V}$ and $A_m' \in \mathbb{R}^{m, m}$; although different in dimension, in practice they represent the same graph.

\begin{theorem}
Let $A, A_m \ge 0$ and irreducible. Then:
\be\label{eqn:bounds}
\sin( P_m\, \mu \,, \tilde{\mu} ) \le
\frac{\gamma}{|| A_m - \lambda \mathbb{1} ||_{2, K_m \cap \tilde{\mu}^{\bot}}} \tan(\mu, K_m)
\ee
where:
 \be\label{eqn:gamma}
 \gamma := || P_m \,A \,(\mathbb{1} - P_m) ||_2 \quad,
 \ee
 \be\label{eqn:tan}
 tan(\mu, K_m) = \sqrt{\frac{\sum_{i \notin \mathcal{I}}\mu_i^2}{\sum_{i \in \mathcal{I}}\mu_i^2}} =  \sqrt{\frac{||\mu||_2^2}{\sum_{i \in \mathcal{I}}\mu_i^2} -1}
 \ee
 
  and $||M||_{2,V}$ is the induced 2 norm of $M$ restricted to the vectorial subspace $V$.
  \end{theorem}
     Equation (\ref{eqn:tan}) is derived by definition of cosine between a vector $\mu$ and a vectorial subspace $K_m$.
  A formal proof can be found in the supplementary (proof 1.1).

The assumptions of having non-negative and irreducible matrices are needed to obtain uniquely defined eigenvalue centralities, as from Perron-Frobenius theorem \cite{golub2012matrix}. In our case, both these hypothesis apply. Except for signed networks, adjacency matrices are always non-negative. In addition, we consider only the strong connected component of the networks considered in our experiments, these have irreducible adjacency matrices.
This theoretical bound motivates the main idea behind our algorithm. We aim at reducing the bound on the sin distance between the real and approximated eigenvalue centrality. Specifically we focus at reducing the factor
$\gamma$. 
Although the sine distance is not the main metric we consider in performance evaluation, we use it as a proxy for the minimization that we are interested in performing. In practice, we observe that this approach works, see section \ref{sec:empirical}. Nonetheless, as said above, the main optimization goal is not on the sine value itself, but on statistics that measure the \textit{relative} rankings of nodes.
In principle, we could reduce the bound in the right-hand side of Eq. (\ref{eqn:bounds}) by minimizing $\tan(\mu, K_m)$ or maximizing $|| A_m - \lambda \mathbb{1} ||_{2, K_m \cap \tilde{\mu}^{\bot}}$. However, these two quantities are difficult to treat analytically and we therefore rely on acting on $\gamma$, which is instead accessible. 
We will partially address the minimization of $\tan(\mu, K_m)$ using theoretically grounded heuristics below.  

 From (\ref{eqn:gamma}), we see that $\gamma$ depends on the choice of the sample. In other words, if we add one node to the sample, the value of $\gamma$ will change. 
 This motivates our proposed sampling algorithm: we consider the \textit{online} addition to the current sample of one node at a time by selecting the best node from the set of non-sampled nodes. The best node is defined as the one whose inclusion in the sample reduces the factor $\gamma$ the most. 
 
  We now explicitly quantify the magnitude of $\gamma$ as a function of adding one candidate node at a time. 
Splitting the adjacency matrix $A = \begin{pmatrix}
A_m' & B \\
C & D
\end{pmatrix} $ we see that $\gamma = || B ||_2$. This is equivalent to the largest singular value $s(B)$ of $B$, i.e. the maximum eigenvalue of the covariance matrix $B^T B$ \cite{golub2012matrix}. This implies that minimizing $\gamma$ is equivalent to minimizing $s(B)$. 
$B$ contains directed edges from outside to inside the sample. Hence, reducing $\gamma$ is related to reducing the number of incoming edges obtained from the sampling. $C$ contains outgoing edges, while $D$ contains edges between nodes outside the sample. While factors not related to edges incoming towards the sample from outside do not appear explicitly in the bound, these are included in the tangent factor \ref{eqn:tan}.

It can be shown, see supplementary section 1.2, that minimizing the entries of $B^T B$ also reduces $s(B)$, hence we now quantify $B^{T}B$. 
Consider a partial sample of size $k$, and a new candidate node to be chosen $j\in \mathcal{V} \setminus \mathcal{I}$. Without loss of generality, we can think of this node as the $k+1$ entry of $A$. This yields the decomposition $B = \begin{pmatrix}
b_1 & U \\
b_2 & b_3^T
\end{pmatrix}$, where $b_1\in \mathbb{R}^{k-1}$ are the edges pointing from $j$ to the nodes already in the subsample, $b_2 \in \mathbb{R}$ is the entry corresponding to $j$, $b_3 \in \mathbb{R}^{n-k+1}$ are edges from nodes outside the sample towards $j$, $U \in \mathbb{R}^{k-1, n- k +1}$ are the edges from nodes outside the sample towards nodes in it, $j$ excluded. We assume $b_2 = 0$, i.e. no self-loops, but the method can be easily generalized to include them. We then greedily pick the node in the graph that, if chosen, minimizes the entries of 
\be \label{eqn:btb}
B^T B = \begin{pmatrix}
b_1^T b_1 & b_1^T U \\
U^T b_1    & U^T U + b_3 b_3^T  
\end{pmatrix}  \quad.
\ee 
This corresponds to maximizing the following quantity:
\begin{equation}\label{th_quantity}
|| b_1 ||_2^2 + || b_1^T U ||_2^2 - || b_3 ||_2^2  \quad.
\end{equation}

Notice that we consider all the terms in \ref{eqn:btb} except $U^T U$. This choice is due to the cost of computing $|| U^T U ||$, since $U^T U$ has size $V$. Give the relatively big size of the matrix, we assume that removing one node will not highly impact its norm, making this part of the optimization negligible compared to the other terms.

We now make few observations. 
Firstly, from the definition of eigenvalue centrality (\ref{eqn:ec}) we aim at minimizing the information missed by not sampling informative \textit{incoming} neighbors of nodes in the sample. We define the \textit{border} of a sampled subgraph as the set of all the incoming neighbors not included in the sample itself. In our online algorithm, we choose to select candidate nodes from those in the border. This choice is arbitrary, but motivated by the following  observations. While selecting candidate nodes as \textit{any} arbitrary node in the network might not be doable in all sampling scenarios \cite{gjoka2010walking}, we assume that accessing neighbors for each node already in the sample is instead possible. This is the case for many real-world applications, e.g. social networks like Twitter and Facebook where follower/followee information is accessible. Second, the effectiveness of this choice is motivated by observing the performance's improvements moving from random nodes sampling (which does not take connecting edges into account) to random walk techniques (which is inherently sampling from neighbors). For comparative results see section \ref{sec:empirical}. Finally, notice that external nodes not in the border have $b_1 = 0$, thus yielding a negative values for the theoretical criterion \ref{th_quantity}. This implies that, in practice, they are unlikely the one chosen when maximizing \ref{th_quantity}. 

\subsection{An extension of the sampling algorithm}
Empirical experiments have shown that addressing the minimization of the bound \ref{eqn:bounds} by considering only $\gamma$ is effective for undirected networks. This doesn't necessarily hold for directed ones. Our guess is that in these cases the term $\tan(\mu, K_m)$ \ref{eqn:tan} also plays an important role in the tightness of the bound in \ref{eqn:bounds}. In order to reduce this quantity, we may then select as candidates, nodes with high centrality. This choice stems from eq \ref{eqn:tan}: the bigger the sum of sampled nodes' centralities, the smaller the tangent term. Since $\mu$ is not available, we approximate it. A possible choice is based on the results of \cite{romance2011local}, suggesting $\mu_j \approx d^{G}_{in}(j)$, the (weighted) in-degree of node $j$. We instead consider $\mu_j \approx d_{in}^{G_k}(j)$, the (weighted) in-degree calculated considering only the incoming edges from nodes that are in the sample. \\
We incorporate this insight into our original criterion (\ref{th_quantity}) by considering the following convex combination:
\be\label{eqn:theor_crit_alpha}
(1 -\alpha) \left(  || b_1 ||_2^2 + || b_1^T U ||_2^2 - || b_3 ||_2^2  \right) + \alpha \,d_{in}^{G_k}(j) \quad,
\ee
where $\alpha$ is an hyperparameter that can be tuned empirically and it allows to interpolate between the two approaches. Using a value of $\alpha$ closer to one means finding nodes with higher degree, therefore aiming to reduce the tangent term. On the other side, setting $\alpha = 0$ means using the vanilla algorithm, only aiming at decreasing the $\gamma$ term.

Empirically, we observe that the size of the border increases like $O(m)$, with a possibly non-negligible constant factor. Thus, at sampling time, we turn to a Monte Carlo approximation: we calculate the value \ref{eqn:theor_crit_alpha} only for a randomly chosen fraction of size $p$ of the new neighbors inside the border. We store the results in a leaderboard for future sampling.
Finally, we impose a maximum size $s$ for the leaderboard, this is for two reasons. First, keeping an ordered leaderboard that allows fast extraction of the maximum element, as well as fast insertion of new ones, is computationally and memory expensive. Secondly, the scores are computed for newly encountered neighbors and stored in the leaderboard for future extraction. The scores given by \ref{eqn:theor_crit_alpha} depend on the current sample. The scores of border nodes that have been computed sometime during past iterations are stored and kept valid for selecting candidate nodes in successive rounds; it is thus advisable to discard old scores by exploiting a small leaderboard size, forcing old nodes to be popped out without being selected when new, potentially higher scoring, nodes are found.

A pseudocode for this is shown in Algorithm \ref{alg:th_sampling}.

\begin{algorithm}[H]%[tb]
   \caption{Theoretical criterion sampling (TCEC)}
   \label{alg:th_sampling}
\begin{algorithmic}
   \STATE {\bfseries Input:} $s$ leaderboard max size, $p$ randomization probability, \\ 
   $m$ final sample size, $1 \le k< m$ random walk initialization size 
    \STATE {\bfseries Output:} $G_m$ final subgraph 
   \STATE - sample $k$ nodes via random walk; call the obtained partial sample $G_k$
   \STATE - store the border of $G_k$ and compute \ref{th_quantity} for a random fraction of its nodes
   \STATE - save these values in a leaderboard
   \STATE - \textbf{while} $size(G_k)< m$:\\
   \qquad pick $v$, highest value neighbor from leaderboard \\
   \qquad $G_{k+1} \leftarrow G_k + v $  \\
   \qquad for a fraction $p$ of the neighbors of $v$, compute \ref{th_quantity} and store in leaderboard
\end{algorithmic}
\end{algorithm}

\subsection{Complexity analysis}
We compare the complexity of our sampling algorithm with the one of random walk methods. At every new node choice, random walk stochastically picks a node among the out-connections of the currently visited one. Random walk has a step cost of $O(1)$. \\
Call $d_{out}, d_{in}$ the average (weighted) out and in degree of the nodes.
Our algorithm performs the following steps: i) extract the best node from the leaderboard; ii) find the new $d_{in}$ neighbors; iii) compute the theoretical criterion for a fraction $p \cdot d_{in}$ of them; iv) store this in the leaderboard. Since a single computation of the theoretical criterion has cost $O(d_{in} + d_{out})$, each step costs
$O(p \cdot d_{in} \cdot (d_{in} + d_{out}))$.
Implementing the leaderboard with a heap with size limit $s$ allows for a cost of $O(1)$ for the extraction of the maximum and a cost of $O(d_{in} \cdot \log(s))$ for the insertion of all new scores. As by design we decide to keep $s$ small, these leaderboard costs are negligible. Finally, notice that with the increase in sample size, the effective number of new neighbors is less than $p \cdot d_{in}$, since many of the neighbors have already been sampled. The final cost is then empirically lower than what expected, and in any case constant with the sample size $k$.

\section{Empirical studies}\label{sec:empirical}

We apply the sampling algorithms to real and synthetic datasets. We describe these datasets below, and we provide a summary of
the datasets in Table \ref{table:datasets}.

\parhead{Synthetic data.} We consider Erd\H{o}s-R\'{e}nyi networks \cite{erdos1959random}, a popular topology when the goal is testing performance on random networks. 

\parhead{Real data.} We use  four real-world open datasets. These datasets include both directed and undirected and both binary and weighted topologies. 

\textit{Internet topology}\footnote{\url{http://irl.cs.ucla.edu/topology/}} \cite{oliveira2008quantifying}: this dataset consists of a a single daily snapshot of a part of the AS (autonomous systems) Internet layer. An \textit{undirected binary}  edge is present between two nodes if a contact between the two has happened during the day.

\textit{Condensed Matter collaboration network\footnote{
\url{https://snap.stanford.edu/data/ca-CondMat.html}}} \cite{leskovec2007graph}:
this dataset contains all the authors in the e-print arXiv Condensed Matter Physics section; an undirected edge is present when two authors collaborated in a publication. 

\textit{Epinions\footnote{
\url{https://snap.stanford.edu/data/soc-Epinions1.html}}} \cite{richardson2003trust}:
this dataset is a type of \textit{who-trusts-whom} directed network extracted from the review site Epionions.com, a consumer reviews website where users rate items and mark users as trustworthy.

\textit{Stanford Web Network}\footnote{\url{https://snap.stanford.edu/data/web-Stanford.html}} \cite{leskovec2009community}:
this dataset represents the pages in the Stanford.edu domain. A \textit{directed binary} edge represents a hyperlink from one page to another.

\parhead{Experimental setup.} For each dataset, we sample subnetworks of different sizes using several sampling algorithms. In addition to ours, we consider mainstream sampling algorithms as random walk \cite{lovasz1993random}, Metropolis-Hastings random walk \cite{metropolis1953equation}, a degree-weighted random walk and uniform sampling on nodes (UniSam). In the results' tables we report only the best performing results among the various random walk sampling techniques listed above and denote it as $RW_{best}$; we denote our method as TCEC. We considered also snowball sampling \cite{goodman1961snowball} and breadth first search (BFS) for comparison. As these provided poorer results then the other methods listed above, we omit them here for the sake of space. Finally, we also include some modern sampling, techniques, such as forest fire \cite{leskovec2006sampling} and expansions sampling \cite{maiya2010sampling}.

\parhead{Implementation details.} Empirically, we find that for undirected graphs the vanilla version (\ref{th_quantity}) of the algorithm, corresponding to $\alpha=0$, leads to good performances; for directed ones we find good performance with $\alpha = 0.5$. All experiments presented here are run using these values. Similarly, the leaderboard maximum size is always set to $s=100$, which we find generally good for most applications. 
We also initialize every sampling round of our algorithm with a random walk exploration of size $1/5$ of the required finaTCECl sample size. This is needed to allow for a degree of reliability in computing the theoretical criterion \ref{eqn:theor_crit_alpha} that requires a minimum sample size to be really discriminative for discerning important nodes. Finally, in order to obtain faster convergence, we set the randomization level $p$ at different values ranging from $0.2$ to $0.8$ depending on the dataset. In practice, $p$ can be chosen after the random walk initialization, when information about the average degree is available.

\parhead{Performance evaluation.} We compare the ground-truth global vector of eigenvalue centrality \textit{restricted} to the entries relative to sampled nodes $P_m \mu$, with the approximate centrality on the sampled network $\tilde{\mu}$. As it is the case for many applications, we are interested in testing whether the \textit{relative} ranking among nodes is preserved, rather than recovering the absolute values of the centrality scores. For this reason we compare the two vectors using the Spearman and Kendall rank correlation coefficients \cite{corder2014nonparametric}. As we empirically see that these two metrics give similar results, we report results only in terms of Kendall and leave the ones with Spearman in the supplementary (section 1.3). One should notice that it is usually relatively easy for any algorithm to obtain high scores on a global scale. This is not surprising, as it is generally easy to tell very different nodes apart, obtaining, for example, a good comparison score in the majority of the possible $\binom{m}{2}$ pairs using Kendall. Things are more difficult on a finer scale, i.e. when selecting nodes nearby in the ranking; in this case nodes are often mistaken in terms of approximate ranking. For this reason we decided to evaluate the two statistics also on a moving window of $10\%$ of the sampled nodes, ordered by global score. We report these results in the supplementary as well (section 1.3).

\begin{table*}[!htp]
  \caption{Datasets Description}
  \label{table:datasets}
  \centering
  \begin{tabular}{cccccc}
    name    & \#nodes   & \#edges   & Average Degree & Directed & Edge Type  \\
    \midrule
    Erd\H{o}s-R\'{e}nyi  & $30000$ & $ 4.5 \cdot 10^6$ & $ 300$ & No & Binary \\
    Internet topology  & $49179$ & $49179$ & $ 3.8 $ & No & Binary \\
    Collaboration Network & $21363$ & $91342$ & $4.3$ & No & Discrete Count \\
    Epinions Network  & $32223$ & $443506$ & $27.5$ & Yes & Binary \\
    Stanford Web Network  & $150532$ & $1576314$ & $20.9$ & Yes & Binary \\
    \bottomrule
  \end{tabular}
\end{table*}

\subsection{Results on synthetic data}
As we can see in tables \ref{table:glob_kendall_erdos} and S1, S2, S3, on an Erd\H{o}s-R\'{e}nyi graph our method performs worse both on a global and local scale. This  is explained by the fact that the fundamental assumption underlying our algorithm, i.e. that some nodes are more important than others for the goodness of estimation, is fundamentally wrong in this case. In fact, in an Erd\H{o}s-R\'{e}nyi graph all nodes share the same statistical properties; this is shown also by the equivalent performances of random walk algorithms and uniform random sampling on nodes in this case. However, it is very unlikely to observe this statistical symmetry in real networks. This is also the only case in which uniform random sampling on nodes performs as well as other methods (see the next sections).

\subsection{Results on real data}
In this case, we can see from Tables \ref{table:glob_kendall_internet} \ref{table:glob_kendall_coauthorship} \ref{table:glob_kendall_epinions} \ref{table:glob_kendall_stanford},  that our algorithm outperforms all other sampling strategies both on a global and local scale in the majority of the dataset. For the \textit{Epinions} network, while the relative trend among other sampling methods is unchanged, ours yields lower performances on smaller sample sizes. Only from sample sizes around $20\%$ our method brings statistical improvements with respect to random walk methods. Our guess is that small sample sizes do not allow a good statistical assessment for the computation of the theoretical criterion \ref{eqn:theor_crit_alpha}, also because in the directed case we expect a heavy tailed distribution of centrality score, which means that detecting important nodes becomes harder due to their small number. Results are presented in table \ref{table:glob_kendall_epinions} (and in supplementary tables S10, S11, S12).
The \textit{Stanford Web Network} is affected by a high level of noise in the results \ref{table:glob_kendall_stanford} (and in supplementary tables S13, S14, S15), signaled by poor performance and negative correlations. This applies to all sampling methods for small sample sizes. Only around $25\%$ sampling size, results start to get more stable and performance improves. On a global scale, statistics grow significantly higher from that point. On the local evaluation window, values are as low as a few percentile points; however, our method obtains statistically significant improvements of around $100\%$ with respect to the best of all random walk sampling algorithms. 

% ------------------------------------------------------------------------------------------------------------------------------------------------------------------------------------
%    ER synthetic
% ------------------------------------------------------------------------------------------------------------------------------------------------------------------------------------
\begin{table}
  \caption{Global Kendall on Erd\H{o}s-R\'{e}nyi Synthetic Graph}
  \label{table:glob_kendall_erdos}
  \centering
  \begin{tabular}{llll}
    $S_{ratio}$    & UniSam   & $RW_{Best}$     &  TCEC \\
    \midrule
    $5\%$ &  $\mathbf{0.145  \pm 0.015} $   & $0.142 \pm 0.016 $  (DW) &  $0.139  \pm 0.020 $ 	\\
    $10\%$ & $\mathbf{0.208 \pm 0.013} $    &  $0.205 \pm 0.014  $  (DW) & $0.165 \pm 0.011 $ 	\\
    $15\%$ &  $0.251 \pm 0.005   $  &   $\mathbf{0.253 \pm 0.009} $  (RW)  & $0.190 \pm 0.009 $	\\
    $20\%$ &  $\mathbf{0.294 \pm 0.007}   $   &  $0.293 \pm 0.009 $ (DW)  & $0.222 \pm 0.007 $	\\
    $25\%$ &    $0.331 \pm 0.004   $  &  $\mathbf{0.334 \pm 0.007} $ (RW)  & $0.246 \pm 0.007 $ 	\\
    $30\%$ &   $0.369 \pm  0.004  $  &  $\mathbf{0.373 \pm 0.006} $ (DW)  & $0.276 \pm 0.007 $	\\
    $40\%$ &   $0.435  \pm 0.004 $  &  $\mathbf{0.436 \pm 0.003} $ (DW)  & $0.345 \pm 0.004 $ 	\\
    \bottomrule
  \end{tabular}
\end{table}

% ------------------------------------------------------------------------------------------------------------------------------------------------------------------------------------
%    Internet topology
% ------------------------------------------------------------------------------------------------------------------------------------------------------------------------------------

\begin{table}
  \caption{Global Kendall on Internet topology Network}
  \label{table:glob_kendall_internet}
  \centering
  \begin{tabular}{llll}
    $S_{ratio}$    & UniSam   & $RW_{Best}$     &  TCEC \\
    \midrule
    $5\%$ &  $0.357  \pm 0.681 $   & $0.977 \pm 0.0022 $  (DW) &  $\mathbf{0.982  \pm 0.0014} $ 	\\
    $10\%$ & $0.303 \pm 0.044 $    &  $0.984 \pm 0.0006  $  (MH) & $\mathbf{0.984 \pm 0.0014} $ 	\\
    $15\%$ &  $0.315 \pm 0.002   $  &   $0.986 \pm 0.0003 $  (MH)  & $\mathbf{0.987 \pm 0.0011} $	\\
    $20\%$ &  $0.328 \pm 0.003   $   &  $0.987 \pm 0.0004 $ (MH)  & $\mathbf{0.990 \pm 0.0002} $	\\
    $25\%$ &    $0.355 \pm 0.003   $  &  $0.989 \pm 0.0003 $ (DW)  & $\mathbf{0.993 \pm 0.0003} $ 	\\
    $30\%$ &   $0.397 \pm  0.001  $  &  $0.990 \pm 0.0002 $ (RW)  & $\mathbf{0.995 \pm 0.0001} $	\\
    $40\%$ &   $0.441  \pm 0.000 $  &  $0.992 \pm 0.0001 $ (MH)  & $\mathbf{0.997 \pm 0.0001} $ 	\\
    \bottomrule
  \end{tabular}
\end{table}

% ------------------------------------------------------------------------------------------------------------------------------------------------------------------------------------
%    Collaboration Network
% ------------------------------------------------------------------------------------------------------------------------------------------------------------------------------------

\begin{table}
  \caption{Global Kendall on Collaboration Network}
  \label{table:glob_kendall_coauthorship}
  \centering
  \begin{tabular}{cccc}
    %\toprule
    %\multicolumn{2}{c}{Part}                   \\
    %\cmidrule(r){1-2}
    $S_{ratio}$    & UniSam   & $RW_{Best}$     &  TCEC \\
        \midrule
    $5\%$ &  $0.245 \pm 0.046 $   & $0.729  \pm 0.064 $  (DW) &  $\mathbf{0.935  \pm 0.003} $ 	\\
    $10\%$ & $0.339 \pm 0.029 $    &  $0.819 \pm 0.032  $  (RW) & $\mathbf{0.958 \pm 0.001} $ 	\\
    $15\%$ &  $0.380 \pm 0.022 $  &   $0.874 \pm 0.007 $  (RW)  & $\mathbf{0.966 \pm 0.001} $	\\
    $20\%$ &  $0.403 \pm 0.015 $   &  $0.900 \pm 0.006 $ (DW)  & $\mathbf{0.970 \pm 0.002} $	\\
    $25\%$ &    $0.451 \pm 0.015 $  &  $0.921 \pm 0.005 $ (DW)  & $\mathbf{0.974 \pm 0.001} $ 	\\
    $30\%$ &   $0.486 \pm 0.014 $  &  $0.938 \pm 0.005 $ (MH)  & $\mathbf{0.977 \pm 0.001} $	\\
    $40\%$ &   $0.545 \pm 0.019 $  &  $0.960 \pm 0.003 $ (MH)  & $\mathbf{0.982 \pm 0.001} $ 	\\
    \bottomrule
  \end{tabular}
\end{table}

% ------------------------------------------------------------------------------------------------------------------------------------------------------------------------------------
%    Epinions Network
% ------------------------------------------------------------------------------------------------------------------------------------------------------------------------------------

\begin{table}
  \caption{Global Kendall on Epinions Network}
  \label{table:glob_kendall_epinions}
  \centering
  \begin{tabular}{cccc}
    %\toprule
    %\multicolumn{2}{c}{Part}                   \\
    %\cmidrule(r){1-2}
    $S_{ratio}$    & UniSam   & $RW_{Best}$     &  TCEC \\    
    \midrule
    $5\%$ &  $0.358  \pm 0.260 $   & $\mathbf{0.634}  \pm 0.010 $  (RW) &  $0.498  \pm 0.009 $ 	\\
    $10\%$ & $0.378 \pm 0.012$    &  $\mathbf{0.649 \pm 0.005}  $  (RW) & $0.601 \pm 0.005 $ 	\\
    $15\%$ &  $0.389 \pm 0.010$  &   $\mathbf{0.660 \pm 0.003} $  (MH)  & $0.658 \pm 0.004 $	\\
    $20\%$ &  $0.409 \pm 0.014$   &  $0.665 \pm 0.004 $ (MH)  & $\mathbf{0.681 \pm 0.004} $	\\
    $25\%$ &    $0.429 \pm 0.017$  &  $0.665 \pm 0.002 $ (RW)  & $\mathbf{0.686 \pm 0.002} $ 	\\
    $30\%$ &   $0.425 \pm 0.007$  &  $0.664 \pm 0.003 $ (DW)  & $\mathbf{0.688 \pm 0.002} $	\\
    $40\%$ &   $0.461 \pm 0.011$  &  $0.657 \pm 0.002 $ (DW)  & $\mathbf{0.676 \pm 0.002} $ 	\\
    \bottomrule
  \end{tabular}
\end{table}

% ------------------------------------------------------------------------------------------------------------------------------------------------------------------------------------
%    Stanford Web  Network
% ------------------------------------------------------------------------------------------------------------------------------------------------------------------------------------

\begin{table}
  \caption{Global Kendall on Stanford Web Network}
  \label{table:glob_kendall_stanford}
  \centering
  \begin{tabular}{cccc}
    %\toprule
    %\multicolumn{2}{c}{Part}                   \\
    %\cmidrule(r){1-2}
    $S_{ratio}$    & UniSam   & $RW_{Best}$     &  TCEC \\    
    \midrule
    $5\%$ &  $-0.350  \pm 0.825 $   & $-0.110  \pm 0.630 $  (MH) &  $\mathbf{0.157  \pm 0.108} $ 	\\
    $10\%$ & $0.064 \pm 0.008$    &  $\mathbf{0.110 \pm 0.010}  $  (DW) & $-0.300 \pm 0.854 $ 	\\
    $15\%$ &  $\mathbf{0.059 \pm 0.009}$  &   $-0.060 \pm 0.648 $  (MH)  & $-0.101 \pm 0.634 $	\\
    $20\%$ &  $0.055 \pm 0.008$   &  $\mathbf{0.164 \pm 0.023} $ (MH)  & $0.092 \pm 0.088 $	\\
    $25\%$ &    $-0.146 \pm 0.618$  &  $0.164 \pm 0.007 $ (RW)  & $\mathbf{0.170 \pm 0.092} $ 	\\
    $30\%$ &   $-0.144 \pm 0.619$  &  $0.161 \pm 0.009 $ (RW)  & $\mathbf{0.258 \pm 0.070} $	\\
    $40\%$ &   $0.078 \pm 0.013$  &  $0.143 \pm 0.007 $ (RW)  & $\mathbf{0.253 \pm 0.018} $ 	\\
    \bottomrule
  \end{tabular}
\end{table}

\section{Conclusions}

While handling network data a relevant problem arises when estimating quantities if one can only access a limited amount of information.
If one has the ability to decide what fraction of the dataset to collect or observe, it is then important to choose the sample wisely in order to recover good estimates of the relevant quantities.
We presented here a novel sampling algorithm for estimating a popular centrality measure on networks, eigenvector centrality. 
The algorithm is theoretically motivated using methods for eigenvalue approximation.
It is computationally scalable and valid for both directed and undirected weighted non-negative adjacency matrices. 

The model relies on lowering the bound on the sin distance between the ground truth and in-sample estimated eigenvector centrality. 
This allows to well preserve the relative ranking of nodes inside the sample ordered by their estimated eigenvector centralities.
It allows to interpolate between quantities related to the network topology which contribute to different extent to the quality of the approximation via hyperparameter tuning. We empirically find that on real networks the vanilla model performs well on undirected networks compared to other standard sampling algorithms; for directed ones performance is improved when using the full model.

Our model assumes that the network topology is static in time. It would be interesting to investigate extensions of the algorithm to the dynamical case, when node or edge removal/addition are possible. 
The method presented in this paper can in principle be adapted for estimating other spectral centrality measures, for instance PageRank.
We focused here in optimizing the estimate of a specific centrality measure. An open question is whether our algorithm, while optimized to preserve this global network property, is indirectly preserving also other network quantities.
While we leave these extensions for future work, we provide an open source implementation of the algorithm\footnote{\url{https://github.com/cdebacco/tcec}}.

%%%%%%%%%% Acknowledgements
%\section*{Acknowledgements}
%\vspace{-0.1in}

%%%%%%%%%% Bibliography
\bibliographystyle{unsrt}
\bibliography{bibliography}

%%%%%%%%%%%%%%%%%%%%%%%%%%%%%%%%%%%%
%%%%%%%%%% SUPPLEMENT STARTS HERE %%%%%%%%%%
%%%%%%%%%%%%%%%%%%%%%%%%%%%%%%%%%%%%

\newcommand{\beginsupplement}{%
        \setcounter{table}{0}
        \renewcommand{\thetable}{S\arabic{table}}%
        \setcounter{figure}{0}
        \renewcommand{\thefigure}{S\arabic{figure}}%
        \setcounter{equation}{0}
        \renewcommand{\theequation}{S\arabic{equation}}
         \setcounter{section}{0}
        \renewcommand{\thesection}{S\arabic{section}}
 }
     
\clearpage
\beginsupplement
\begin{widetext}

%\part*{\titlefont{Supporting Information (SI)}}
\section*{{Supporting Information (SI)}}

\section{Additional Theorems and Proofs}
\subsection{Proof of theorem 1}
\begin{proof}
We follow the same notation of the proof in \cite[p.103/104]{saad}.
What we derive here is a specialization of their proof to our case. \\
Call $\phi$ the angle between $\mu$ and $K_m$ and $\omega$ the angle between $P_m \mu$ and $\tilde{\mu}$. Therefore we can decompose 
$$ \mu = v \cos \phi + w \sin \phi $$
where $w, v$ are the normalized projections of $\mu$ respectively on $K_m$ and $K_m^{\bot}$. Multiplying both sides by $(A-\lambda \mathbb{1})$, since $A\mu = \lambda\mu$, we obtain
$$ (A - \lambda \mathbb{1}) v \cos \phi + (A - \lambda \mathbb{1}) w \sin \phi = 0 $$ 
By multiplying for $P_m$ and taking the norm we obtain 
\begin{equation}\label{bound_proof}
|| P_m (A - \lambda \mathbb{1}) v || \cos \phi = 
|| P_m (A - \lambda \mathbb{1}) w || \sin \phi 
\le \gamma \sin \phi
\end{equation} 
where the second passage derives from the fact that $w = (\mathbb{1} - P_m) w$ therefore
\begin{align*}
|| P_m (A - \lambda \mathbb{1}) w || &= 
|| P_m (A - \lambda \mathbb{1}) (\mathbb{1} - P_m) w || \\ 
&= || P_m A (\mathbb{1} - P_m) w - \lambda P_m  (\mathbb{1} - P_m) w|| \\
&= || P_m A (\mathbb{1} - P_m) w || \\
& \le || P_m A (\mathbb{1} - P_m) || = \gamma
\end{align*}
Now, consider the vector $z \in K_m, z \bot \tilde{\mu}$ such that
$$ v = \tilde{\mu} \cos \omega + z \sin \omega $$
Then applying the left multiplication $P_m (A - \lambda\mathbb{1})$ we obtain
$$ P_m (A - \lambda\mathbb{1}) v = (\tilde{\lambda} - \lambda) \tilde{\mu} \cos \omega + P_m (A - \lambda\mathbb{1})z \sin \omega $$
which implies 
\begin{align*}
||P_m (A - \lambda\mathbb{1}) v|| 
&\ge || P_m (A - \lambda\mathbb{1})z || \sin \omega \\
&= || (P_m A P_m - \lambda\mathbb{1})z || \sin \omega \\
&= || (A_m - \lambda \mathbb{1}) ||_{2, K_m \cap \tilde{\mu}^{\bot}} \sin \omega
\end{align*}
Then using \eqref{bound_proof} we obtain
$$ || (A_m - \lambda \mathbb{1}) ||_{2, K_m \cap \tilde{\mu}^{\bot}} \sin \omega \cos \phi \le \gamma \sin \phi  $$
which concludes the proof since $\sin \omega = \sin(P_m \mu, \tilde{\mu})$ and $\sin \phi = \sin (\mu, K_m)$
\end{proof}

\subsection{Additional Theorems}

\begin{lemma}
Let $0 \le B \in \mathbb{R}^{s,t}$ (where $0 \le B$ is intended elementwise, i.e. $0 \le B_{i,j} \forall i,j$). Then the largest singular value of $B$, $s(B)$, satisifies the following
\begin{align*}
s(B) &= \max_{||x||=1} x^T B^T Bx = \max_{||x||=1} ||B x||_2^2 \\
&=  \max_{||x||=1, x \ge 0} ||B x||_2^2 = \max_{||x||=1, x \ge 0} x^T B^T B x
\end{align*} 
where the first equality derives from Courant-Fischer theorem \cite[p24]{saad}, as $s(B)$ is the biggest eigenvalue of $B^T B$, and the third one from the fact that $B \ge 0$
\end{lemma}

\begin{theorem}\label{th:singular_val}
Let $0 \le B, C \in \mathbb{R}^{s,t}$, $B^T B \le C^T C$. Then $s(B) \le s(C)$. (This theorem holds the same if we only suppose $0 \le B \le C$, but we use this version for developing the algorithm)
\begin{proof}
Using the previous lemma
$$ s(B) =\max_{||x||=1, x \ge 0} x^T B^T B x
 \le\max_{||x||=1, x \ge 0} x^T C^T C x = s(C) $$
(to prove in the case $B \le C$, just use 
$ s(B) =\max_{||x||=1, x \ge 0} ||B x||_2^2
 \le \max_{||x||=1, x \ge 0} ||C x||_2^2 = s(C) $
)
\end{proof}
\end{theorem}

%%%%%%%%%% TABLES %%%%%%%%%%
\clearpage
\section{Supplemental Tables}

In this section we provide the results that are not included in the original paper. These are made of the Spearman score both on the global score vectors and on a moving window of size $10\%$ of the entire sample. The moving window scores for the Kendall Tau statistic are reported here as well. Notice that in some of the columns corresponding to the moving window score for the uniform sampling, some NA entries are present. These are placeholders to indicate where it hasn't been possible to perform the computations due numerical reasons.

% ------------------------------------------------------------------------------------------------------------------------------------------------------------------------------------
%    ER synthetic
% ------------------------------------------------------------------------------------------------------------------------------------------------------------------------------------

\begin{table}[htb]
  \caption{Global Spearman on Erd\H{o}s-R\'{e}nyi Synthetic Graph}
  \label{table:glob_spearman_erdos}
  \centering
  \begin{tabular}{cccc}
    %\toprule
    %\multicolumn{2}{c}{Part}                   \\
    %\cmidrule(r){1-2}
    $S_{ratio}$    & UniSam   & $RW_{Best}$     &  TCEC \\
        \midrule
    $5\%$ &  $0.103  \pm 0.023 $   & $\mathbf{0.211 \pm 0.024} $  (DW) &  $0.206  \pm 0.029 $ 	\\
    $10\%$ & $0.169 \pm 0.018 $    &  $\mathbf{0.304 \pm 0.020 } $  (DW) & $0.245 \pm 0.017 $ 	\\
    $15\%$ &  $0.251 \pm 0.007   $  &   $\mathbf{0.371 \pm 0.012} $  (RW)  & $0.282 \pm 0.014 $	\\
    $20\%$ &  $0.265 \pm 0.009   $   &  $0.293 \pm 0.013 $ (DW)  & $\mathbf{0.327 \pm 0.010} $	\\
    $25\%$ &    $0.331 \pm 0.006   $  &  $\mathbf{0.484 \pm 0.010} $ (RW)  & $0.361 \pm 0.011 $ 	\\
    $30\%$ &   $0.369 \pm  0.005  $  &  $\mathbf{0.529 \pm 0.007} $ (RW)  & $0.404 \pm 0.010 $	\\
    $40\%$ &   $0.435  \pm 0.005 $  &  $\mathbf{0.612 \pm 0.006} $ (MH)  & $0.500 \pm 0.005 $ 	\\
    \bottomrule
  \end{tabular}
\end{table}

\begin{table}[htb]
  \caption{Moving Window Spearman on Erd\H{o}s-R\'{e}nyi Synthetic Graph}
  \label{table:local_spearman_erdos}
  \centering
  \begin{tabular}{cccc}
    %\toprule
    %\multicolumn{2}{c}{Part}                   \\
    %\cmidrule(r){1-2}
    $S_{ratio}$    & UniSam   & $RW_{Best}$     &  TCEC \\
        \midrule
    $5\%$ &  $\mathbf{0.026  \pm 0.003} $   & $0.025 \pm 0.006 $  (MH) &  $\mathbf{0.026  \pm 0.004} $ 	\\
    $10\%$ & $\mathbf{0.036 \pm 0.003} $    &  $\mathbf{0.036 \pm 0.002 } $  (MH) & $0.029 \pm 0.003 $ 	\\
    $15\%$ &  $0.044 \pm 0.002   $  &   $\mathbf{0.045 \pm 0.003} $  (RW)  & $0.035 \pm 0.003 $	\\
    $20\%$ &  $\mathbf{0.053 \pm 0.002}   $   &  $\mathbf{0.053 \pm 0.003} $ (DW)  & $0.039 \pm 0.002 $	\\
    $25\%$ &    $0.061 \pm 0.002   $  &  $\mathbf{0.062 \pm 0.002} $ (RW)  & $0.045 \pm 0.001 $ 	\\
    $30\%$ &   $0.069 \pm  0.002  $  &  $\mathbf{0.070 \pm 0.002} $ (DW)  & $0.050 \pm 0.001 $	\\
    $40\%$ &   $\mathbf{0.087  \pm 0.002} $  &  $\mathbf{0.087 \pm 0.001} $ (DW)  & $0.066 \pm 0.001 $ 	\\
    \bottomrule
  \end{tabular}
\end{table}

\begin{table}[htb]
  \caption{Moving Window Kendall on Erd\H{o}s-R\'{e}nyi Synthetic Graph}
  \label{table:local_kendall_erdos}
  \centering
  \begin{tabular}{llll}
    %\toprule
    %\multicolumn{2}{c}{Part}                   \\
    %\cmidrule(r){1-2}
    $S_{ratio}$    & UniSam   & $RW_{Best}$     &  TCEC \\
    \midrule
    $5\%$ &  $\mathbf{0.018  \pm 0.002} $   & $0.017 \pm 0.004 $  (MH) &  $0.017  \pm 0.002 $ 	\\
    $10\%$ & $\mathbf{0.024 \pm 0.002} $    &  $\mathbf{0.024 \pm 0.001}  $  (MH) & $0.019 \pm 0.002 $ 	\\
    $15\%$ &  $0.029 \pm 0.001   $  &   $\mathbf{0.030 \pm 0.002} $  (RW)  & $0.023 \pm 0.002 $	\\
    $20\%$ &  $\mathbf{0.036 \pm 0.001}   $   &  $0.035 \pm 0.002 $ (DW)  & $0.026 \pm 0.001 $	\\
    $25\%$ &    $\mathbf{0.041 \pm 0.001}   $  &  $\mathbf{0.041 \pm 0.001} $ (MH)  & $0.030 \pm 0.001 $ 	\\
    $30\%$ &   $0.046 \pm  0.001  $  &  $\mathbf{0.047 \pm 0.001} $ (DW)  & $0.033 \pm 0.001 $	\\
    $40\%$ &   $\mathbf{0.058  \pm 0.001} $  &  $\mathbf{0.058 \pm 0.000} $ (DW)  & $0.044 \pm 0.001 $ 	\\
    \bottomrule
  \end{tabular}
\end{table}

% ------------------------------------------------------------------------------------------------------------------------------------------------------------------------------------
%    Internet topology
% ------------------------------------------------------------------------------------------------------------------------------------------------------------------------------------

\begin{table}[htb]
  \caption{Global Spearman on Internet Topology Network}
  \label{table:glob_spearman_internet}
  \centering
  \begin{tabular}{cccc}
    %\toprule
    %\multicolumn{2}{c}{Part}                   \\
    %\cmidrule(r){1-2}
    $S_{ratio}$    & UniSam   & $RW_{Best}$     &  TCEC \\
        \midrule
    $5\%$ &  $0.095  \pm 0.702 $   & $\mathbf{0.999 \pm 0.000} $  (DW) &  $\mathbf{0.999  \pm 0.000} $ 	\\
    $10\%$ & $0.388 \pm 0.057 $    &  $\mathbf{0.999 \pm 0.000 } $  (MH) & $\mathbf{0.999 \pm 0.000} $ 	\\
    $15\%$ &  $0.402 \pm 0.072   $  &   $\mathbf{0.999 \pm 0.000} $  (MH)  & $\mathbf{0.999 \pm 0.000} $	\\
    $20\%$ &  $0.422 \pm 0.109   $   &  $\mathbf{0.999 \pm 0.000} $ (DW)  & $\mathbf{0.999 \pm 0.000} $	\\
    $25\%$ &    $0.454 \pm 0.057   $  &  $\mathbf{0.999 \pm 0.000} $ (DW)  & $\mathbf{0.999 \pm 0.000} $ 	\\
    $30\%$ &   $0.502 \pm  0.077  $  &  $\mathbf{0.999 \pm 0.000} $ (DW)  & $\mathbf{0.999 \pm 0.000} $	\\
    $40\%$ &   $0.552  \pm 0.094 $  &  $\mathbf{0.999 \pm 0.000} $ (MH)  & $\mathbf{0.999 \pm 0.000} $ 	\\
    \bottomrule
  \end{tabular}
\end{table}

\begin{table}[htb]
  \caption{Moving Window Spearman on Internet Topology Network}
  \label{table:local_spearman_internet}
  \centering
  \begin{tabular}{cccc}
    %\toprule
    %\multicolumn{2}{c}{Part}                   \\
    %\cmidrule(r){1-2}
    $S_{ratio}$    & UniSam   & $RW_{Best}$     &  TCEC \\
        \midrule
    $5\%$ &  NA   & $0.928 \pm 0.011 $  (RW) &  $\mathbf{0.950  \pm 0.006} $ 	\\
    $10\%$ &  NA    &  $0.955 \pm 0.003  $  (MH) & $\mathbf{0.958 \pm 0.006} $ 	\\
    $15\%$ &  NA  &   $0.965 \pm 0.002 $  (DW)  & $\mathbf{0.969 \pm 0.004} $	\\
    $20\%$ &  NA     &  $0.971 \pm 0.002 $ (DW)  & $\mathbf{0.980 \pm 0.001} $	\\
    $25\%$ &    NA  &  $0.976 \pm 0.001 $ (DW)  & $\mathbf{0.989 \pm 0.001} $ 	\\
    $30\%$ &   NA &  $0.980 \pm 0.001 $ (DW)  & $\mathbf{0.993 \pm 0.001} $	\\
    $40\%$ &   NA  &  $0.987 \pm 0.000 $ (RW)  & $\mathbf{0.997 \pm 0.000} $ 	\\
    \bottomrule
  \end{tabular}
\end{table}

\begin{table}[htb]
  \caption{Moving Window Kendall on Internet topology Network}
  \label{table:local_kendall_internet}
  \centering
  \begin{tabular}{cccc}
    %\toprule
    %\multicolumn{2}{c}{Part}                   \\
    %\cmidrule(r){1-2}
    $S_{ratio}$    & UniSam   & $RW_{Best}$     &  TCEC \\    
    \midrule
    $5\%$ &  NA   & $0.810 \pm 0.014 $  (RW) &  $\mathbf{0.839  \pm 0.012} $ 	\\
    $10\%$ & NA    &  $0.852 \pm 0.005  $  (MH) & $\mathbf{0.854 \pm 0.011} $ 	\\
    $15\%$ &  NA  &   $0.869 \pm 0.004 $  (DW)  & $\mathbf{0.880 \pm 0.010} $	\\
    $20\%$ &  NA  &  $0.880 \pm 0.003 $ (DW)  & $\mathbf{0.907 \pm 0.002} $	\\
    $25\%$ &    NA &  $0.892 \pm 0.002 $ (DW)  & $\mathbf{0.934 \pm 0.004} $ 	\\
    $30\%$ &   NA  &  $0.903 \pm 0.002 $ (RW)  & $\mathbf{0.952 \pm 0.002} $	\\
    $40\%$ &   NA  &  $0.923 \pm 0.001 $ (MH)  & $\mathbf{0.974 \pm 0.001} $ 	\\
    \bottomrule
  \end{tabular}
\end{table}

% ------------------------------------------------------------------------------------------------------------------------------------------------------------------------------------
%    Collaboration Network
% ------------------------------------------------------------------------------------------------------------------------------------------------------------------------------------

\begin{table}[htb]
  \caption{Global Spearman on Collaboration Network}
  \label{table:glob_spearman_coauthorship}
  \centering
  \begin{tabular}{cccc}
    %\toprule
    %\multicolumn{2}{c}{Part}                   \\
    %\cmidrule(r){1-2}
    $S_{ratio}$    & UniSam   & $RW_{Best}$     &  TCEC \\
        \midrule
    $5\%$ &  NA  & $0.893 \pm 0.050 $  (DW) &  $\mathbf{0.994  \pm 0.001} $ 	\\
    $10\%$ & NA    &  $0.950 \pm 0.016  $  (RW) & $\mathbf{0.997 \pm 0.000} $ 	\\
    $15\%$ &  NA  &   $0.974 \pm 0.003 $  (RW)  & $\mathbf{0.998 \pm 0.000} $	\\
    $20\%$ &  NA   &  $0.982 \pm 0.002 $ (DW)  & $\mathbf{0.998 \pm 0.000} $	\\
    $25\%$ &    NA  &  $0.988 \pm 0.002 $ (MH)  & $\mathbf{0.999 \pm 0.000} $ 	\\
    $30\%$ &   NA &  $0.992 \pm 0.001 $ (MH)  & $\mathbf{0.999 \pm 0.000} $	\\
    $40\%$ &   NA  &  $0.996 \pm 0.001 $ (MH)  & $\mathbf{0.999 \pm 0.000} $ 	\\
    \bottomrule
  \end{tabular}
\end{table}

\begin{table}[htb]
  \caption{Moving Window Spearman on Collaboration Network}
  \label{table:local_spearman_coauthorship}
  \centering
  \begin{tabular}{cccc}
    %\toprule
    %\multicolumn{2}{c}{Part}                   \\
    %\cmidrule(r){1-2}
    $S_{ratio}$    & UniSam   & $RW_{Best}$     &  TCEC \\
        \midrule
    $5\%$ &  NA   & $0.250 \pm 0.063 $  (DW) &  $\mathbf{0.692  \pm 0.019} $ 	\\
    $10\%$ & NA    &  $0.370 \pm 0.055  $  (RW) & $\mathbf{0.826 \pm 0.007} $ 	\\
    $15\%$ &  NA &   $0.503 \pm 0.024 $  (RW)  & $\mathbf{0.875 \pm 0.008} $	\\
    $20\%$ &  NA  &  $0.593 \pm 0.018 $ (RW)  & $\mathbf{0.899 \pm 0.007} $	\\
    $25\%$ &    NA  &  $0.677 \pm 0.025 $ (DW)  & $\mathbf{0.920 \pm 0.006} $ 	\\
    $30\%$ &   NA  &  $0.752 \pm 0.015 $ (DW)  & $\mathbf{0.932 \pm 0.004} $	\\
    $40\%$ &   NA &  $0.855 \pm 0.013 $ (MH)  & $\mathbf{0.950 \pm 0.002} $ 	\\
    \bottomrule
  \end{tabular}
\end{table}

\begin{table}[htb]
  \caption{Moving Window Kendall on Collaboration Network}
  \label{table:local_kendall_coauthorship}
  \centering

   \begin{tabular}{cccc}
    %\toprule
    %\multicolumn{2}{c}{Part}                   \\
    %\cmidrule(r){1-2}
    $S_{ratio}$    & UniSam   & $RW_{Best}$     &  TCEC \\
      \midrule
    $5\%$ &  NA  & $0.729  \pm 0.064 $  (DW) &  $\mathbf{0.935  \pm 0.003} $ 	\\
    $10\%$ & NA    &  $0.819 \pm 0.032  $  (RW) & $\mathbf{0.958 \pm 0.001} $ 	\\
    $15\%$ &  NA  &   $0.874 \pm 0.007 $  (RW)  & $\mathbf{0.966 \pm 0.001} $	\\
    $20\%$ &  NA   &  $0.900 \pm 0.006 $ (DW)  & $\mathbf{0.970 \pm 0.002} $	\\
    $25\%$ &    NA  &  $0.921 \pm 0.005 $ (DW)  & $\mathbf{0.974 \pm 0.001} $ 	\\
    $30\%$ &   NA  &  $0.938 \pm 0.005 $ (MH)  & $\mathbf{0.977 \pm 0.001} $	\\
    $40\%$ &    NA  &  $0.960 \pm 0.003 $ (MH)  & $\mathbf{0.982 \pm 0.001} $ 	\\
    \bottomrule
  \end{tabular}

\end{table}

% ------------------------------------------------------------------------------------------------------------------------------------------------------------------------------------
%    Epinions Network
% ------------------------------------------------------------------------------------------------------------------------------------------------------------------------------------

\begin{table}[htb]
  \caption{Global Spearman on Epinions Network}
  \label{table:glob_spearman_epinions}
  \centering
  \begin{tabular}{cccc}
    %\toprule
    %\multicolumn{2}{c}{Part}                   \\
    %\cmidrule(r){1-2}
    $S_{ratio}$    & UniSam   & $RW_{Best}$     &  TCEC \\
        \midrule
    $5\%$ &  $0.455 \pm 0.033 $   & $\mathbf{0.822 \pm 0.009 }$  (RW) &  $0.681  \pm 0.011 $ 	\\
    $10\%$ & $0.490 \pm 0.015  $    &  $\mathbf{0.840 \pm 0.004}  $  (RW) & $0.792 \pm 0.005 $ 	\\
    $15\%$ &  $0.510 \pm 0.013 $  &   $\mathbf{0.852 \pm 0.003} $  (MH)  & $0.847 \pm 0.004 $	\\
    $20\%$ &  $0.541 \pm 0.018 $   &  $0.856 \pm 0.003 $ (MH)  & $\mathbf{0.868 \pm 0.003} $	\\
    $25\%$ &    $0.571 \pm 0.023  $  &  $0.856 \pm 0.002 $ (RW)  & $\mathbf{0.873 \pm 0.002} $ 	\\
    $30\%$ &   $0.571 \pm 0.006 $  &  $0.854 \pm 0.001 $ (MH)  & $\mathbf{0.874 \pm 0.002} $	\\
    $40\%$ &   $0.619 \pm 0.015  $  &  $0.856 \pm 0.001 $ (RW)  & $\mathbf{0.862 \pm 0.002} $ 	\\
    \bottomrule
  \end{tabular}
\end{table}

\begin{table}[htb]
  \caption{Moving Window Spearman on Epinions Network}
  \label{table:local_spearman_epinions}
  \centering
  \begin{tabular}{cccc}
    %\toprule
    %\multicolumn{2}{c}{Part}                   \\
    %\cmidrule(r){1-2}
    $S_{ratio}$    & UniSam   & $RW_{Best}$     &  TCEC \\
        \midrule
    $5\%$ &  NA   & $\mathbf{0.168 \pm 0.007}$  (RW) &  $0.119  \pm 0.003 $ 	\\
    $10\%$ & NA    &  $\mathbf{0.170 \pm 0.002}  $  (DW) & $0.162 \pm 0.004 $ 	\\
    $15\%$ &  NA  &   $0.173 \pm 0.003 $  (DW)  & $\mathbf{0.181 \pm 0.003} $	\\
    $20\%$ &  NA   &  $0.174 \pm 0.001 $ (MH)  & $\mathbf{0.188 \pm 0.002} $	\\
    $25\%$ &    NA  &  $0.174 \pm 0.002 $ (RW)  & $\mathbf{0.188 \pm 0.002} $ 	\\
    $30\%$ &   NA  &  $0.175 \pm 0.002 $ (DW)  & $\mathbf{0.188 \pm 0.001} $	\\
    $40\%$ &   NA  &  $0.176 \pm 0.002 $ (DW)  & $\mathbf{0.185 \pm 0.001} $ 	\\
    \bottomrule
  \end{tabular}
\end{table}

\begin{table}[htb]
  \caption{Moving Window Kendall on Epinions Network}
  \label{table:local_kendall_epinions}
  \centering
  \begin{tabular}{cccc}
    %\toprule
    %\multicolumn{2}{c}{Part}                   \\
    %\cmidrule(r){1-2}
    $S_{ratio}$    & UniSam   & $RW_{Best}$     &  TCEC \\    
    \midrule
    $5\%$ &  NA   & $\mathbf{0.114}  \pm 0.005 $  (RW) &  $0.081  \pm 0.002 $ 	\\
    $10\%$ & NA    &  $\mathbf{0.115 \pm 0.001}  $  (DW) & $0.109 \pm 0.003 $ 	\\
    $15\%$ &  NA  &   $\mathbf{0.116 \pm 0.002} $  (DW)  & $0.122 \pm 0.002 $	\\
    $20\%$ &  NA   &  $0.117 \pm 0.002 $ (MH)  & $\mathbf{0.127 \pm 0.002} $	\\
    $25\%$ &    NA &  $0.117 \pm 0.001 $ (RW)  & $\mathbf{0.126 \pm 0.001} $ 	\\
    $30\%$ &   NA  &  $0.118 \pm 0.001 $ (DW)  & $\mathbf{0.127 \pm 0.001} $	\\
    $40\%$ &   NA  &  $0.118 \pm 0.001 $ (DW)  & $\mathbf{0.124 \pm 0.001} $ 	\\
    \bottomrule
  \end{tabular}
\end{table}

% ------------------------------------------------------------------------------------------------------------------------------------------------------------------------------------
%    Stanford Web Network
% ------------------------------------------------------------------------------------------------------------------------------------------------------------------------------------

\begin{table}[htb]
  \caption{Global Spearman on Stanford Web Network}
  \label{table:glob_spearman_stanford}
  \centering
  \begin{tabular}{cccc}
    %\toprule
    %\multicolumn{2}{c}{Part}                   \\
    %\cmidrule(r){1-2}
    $S_{ratio}$    & UniSam   & $RW_{Best}$     &  TCEC \\
        \midrule
    $5\%$ &  $-0.336 \pm 0.832 $   & $\mathbf{-0.064 \pm 0.646 }$  (MH) &  $0.228  \pm 0.156 $ 	\\
    $10\%$ & $0.083 \pm 0.011  $    &  $\mathbf{0.167 \pm 0.015}  $  (DW) & $-0.254 \pm 0.881 $ 	\\
    $15\%$ &  $\mathbf{0.079 \pm 0.012} $  &   $0.006 \pm 0.671 $  (MH)  & $-0.055 \pm 0.654 $	\\
    $20\%$ &  $0.075 \pm 0.010 $   &  $\mathbf{0.240 \pm 0.031} $ (MH)  & $0.127 \pm 0.125 $	\\
    $25\%$ &    $-0.125 \pm 0.625  $  &  $\mathbf{0.239 \pm 0.010} $ (RW)  & $0.234 \pm 0.132 $ 	\\
    $30\%$ &   $-0.122 \pm 0.626 $  &  $0.235 \pm 0.012 $ (RW)  & $\mathbf{0.365 \pm 0.097} $	\\
    $40\%$ &   $0.110 \pm 0.019  $  &  $0.209 \pm 0.010 $ (RW)  & $\mathbf{0.355 \pm 0.025} $ 	\\
    \bottomrule
  \end{tabular}
\end{table}

\begin{table}[htb]
  \caption{Moving Window Spearman on Stanford Web Network}
  \label{table:local_spearman_stanford}
  \centering
  \begin{tabular}{cccc}
    %\toprule
    %\multicolumn{2}{c}{Part}                   \\
    %\cmidrule(r){1-2}
    $S_{ratio}$    & UniSam   & $RW_{Best}$     &  TCEC \\
        \midrule
    $5\%$ &    NA   & $-0.018 \pm 0.606$  (MH) &  $\mathbf{0.005  \pm 0.030 }$ 	\\
    $10\%$ & NA    &  $\mathbf{0.020 \pm 0.003}  $  (DW) & $-0.372 \pm 0.815 $ 	\\
    $15\%$ &  NA  &   $-0.170 \pm 0.610 $  (MH)  & $\mathbf{-0.167 \pm 0.611} $	\\
    $20\%$ &  NA   &  $0.035 \pm 0.006 $ (MH)  & $\mathbf{0.054 \pm 0.028} $	\\
    $25\%$ &    NA  &  $0.037 \pm 0.004 $ (MH)  & $\mathbf{0.064 \pm 0.021} $ 	\\
    $30\%$ &   NA  &  $0.037 \pm 0.002 $ (RW)  & $\mathbf{0.067 \pm 0.016} $	\\
    $40\%$ &   NA  &  $0.032 \pm 0.001 $ (RW)  & $\mathbf{0.062 \pm 0.005} $ 	\\
    \bottomrule
  \end{tabular}
\end{table}

\begin{table}[htb]
  \caption{Moving window Kendall on Stanford Web Network}
  \label{table:local_kendall_stanford}
  \centering
  \begin{tabular}{cccc}
    %\toprule
    %\multicolumn{2}{c}{Part}                   \\
    %\cmidrule(r){1-2}
    $S_{ratio}$    & UniSam   & $RW_{Best}$     &  TCEC \\    
    \midrule
    $5\%$ &  NA   & $-0.187  \pm 0.604 $  (MH) &  $\mathbf{0.035  \pm 0.020} $ 	\\
    $10\%$ & NA    &  $\mathbf{0.014 \pm 0.002}  $  (DW) & $-0.381 \pm 0.810 $ 	\\
    $15\%$ &  NA  &   $-0.179 \pm 0.607 $  (MH)  & $\mathbf{-0.178 \pm 0.608} $	\\
    $20\%$ &  NA   &  $0.024 \pm 0.004 $ (MH)  & $\mathbf{0.037 \pm 0.019} $	\\
    $25\%$ &    NA  &  $0.025 \pm 0.003 $ (MH)  & $\mathbf{0.044 \pm 0.015} $ 	\\
    $30\%$ &   NA  &  $0.025 \pm 0.001 $ (RW)  & $\mathbf{0.046 \pm 0.011} $	\\
    $40\%$ &   NA  &  $0.022 \pm 0.001 $ (RW)  & $\mathbf{0.043 \pm 0.004} $ 	\\
    \bottomrule
  \end{tabular}
\end{table}

\end{widetext}

\end{document}